\documentclass[fleqn,usenatbib]{mnras}

\usepackage{newtxtext,newtxmath}
\usepackage{anyfontsize}

\usepackage[T1]{fontenc}

\DeclareRobustCommand{\VAN}[3]{#2}
\let\VANthebibliography\thebibliography
\def\thebibliography{\DeclareRobustCommand{\VAN}[3]{##3}\VANthebibliography}

\usepackage{graphicx}	
\usepackage{amsmath}	
\usepackage{orcidlink}

\newcommand{\afflink}[1]{\hyperlink{aff:#1}{\textsuperscript{#1}}}

\title[Following up the \textit{Kepler} field with \textit{PLATO}]{Following up the \textit{Kepler} field with \textit{PLATO}: Transit Timing Performance}

\author[M. A. Mitchell et al.]{Morgan A. Mitchell$^{\orcidlink{0009-0004-6130-7775}}$,\afflink{1}$^,$\afflink{2}\thanks{E-mail: morgan\_mitchell@outlook.com}
James McCormac$^{\orcidlink{0000-0003-1631-4170}}$,\afflink{1}$^,$\afflink{2}$^,$\afflink{3}
Don Pollacco$^{\orcidlink{0000-0001-9850-9697}}$,\afflink{1}$^,$\afflink{2}$^,$\afflink{3}
Emmanuel Grolleau$^{\orcidlink{0000-0002-6453-8737}}$,\afflink{4}
\newauthor
Nicholas Jannsen$^{\orcidlink{0000-0003-4670-9616}}$,\afflink{5}
Daniel R. Reese$^{\orcidlink{0000-0003-4854-7550}}$,\afflink{4}
Réza Samadi$^{\orcidlink{0000-0003-1446-8934}}$,\afflink{4}
Yoshi Nike Emilia Eschen$^{\orcidlink{0009-0006-6397-2503}}$,\afflink{1}$^,$\afflink{2}$^,$\afflink{6}
\newauthor
Ioannis Apergis$^{\orcidlink{0009-0004-7473-4573}}$,\afflink{1}$^,$\afflink{2}
James A. Blake$^{\orcidlink{0000-0002-5903-2387}}$,\afflink{1}$^,$\afflink{3} 
David J. A. Brown$^{\orcidlink{0000-0003-1098-2442}}$,\afflink{1}$^,$\afflink{2}
Lauren Doyle$^{\orcidlink{0000-0002-9365-2555}}$,\afflink{1}$^,$\afflink{2}
\newauthor
Isobel S. Lockley$^{\orcidlink{0009-0003-0928-3588}}$\afflink{1}$^,$\afflink{2} 
and
Lixian Shen$^{\orcidlink{0009-0003-3618-4841}}$\afflink{1}$^,$\afflink{2}
\\
\hypertarget{aff:1}$^{1}$Department of Physics, University of Warwick, Gibbet Hill Road, Coventry CV4 7AL, UK\\
\hypertarget{aff:2}$^{2}$Centre for Exoplanets and Habitability, University of Warwick, Gibbet Hill Road, Coventry CV4 7AL, UK\\
\hypertarget{aff:3}$^{3}$Centre for Space Domain Awareness, University of Warwick, Gibbet Hill Road, Coventry CV4 7AL, UK\\
\hypertarget{aff:4}$^{4}$LIRA, Observatoire de Paris, Université PSL, Sorbonne Université, Université Paris Cité, CY Cergy Paris Université, CNRS, 92190 Meudon, France\\
\hypertarget{aff:5}$^{5}$Institute for Astronomy, KU Leuven, Celestijnenlaan 200D bus 2401, 3001 Leuven, Belgium\\
\hypertarget{aff:6}$^{6}$Isaac Newton Group of Telescopes, Apartado de Correos 321, E-38700, Santa Cruz de la Palma, Spain\\
}

\date{Accepted XXX. Received YYY; in original form ZZZ}

\pubyear{2026}

\begin{document}
\label{firstpage}
\pagerange{\pageref{firstpage}--\pageref{lastpage}}
\maketitle

\begin{abstract}
The European Space Agency is set to launch \textit{PLATO}, the third medium-class mission of its Cosmic Vision programme, in early 2027. Using the transit method, \textit{PLATO} is expected to detect thousands of exoplanets orbiting bright, nearby stars of spectral types F5--K7. Although the mission is primarily designed to enable mass measurements via radial velocities, its precise photometry and long observational baselines may also permit the detection of transit timing variations (TTVs), which can provide complementary dynamical constraints in multi-planet systems. One possible \textit{PLATO} observing scenario involves a two-year-long observation of a Northern field that may partially or fully overlap with the original \textit{Kepler} field, creating an opportunity to revisit known multi-planet systems with a photometric baseline exceeding 20 years. We simulate \textit{PLATO} observations of 152 \textit{Kepler} host stars containing at least one planet with previously detected TTVs, yielding a sample of 361 transiting planets. Our CCD-level simulations incorporate realistic stellar variability and employ both aperture and point spread function (PSF)-fitting photometry, accounting for each target’s real photometric contaminants. While the extended temporal baseline offers the potential for improved dynamical constraints in favourable cases, our simulations show that this potential is strongest for carefully selected systems, as \textit{PLATO}'s smaller collecting area and larger pixel scale limit the achievable per-transit precision relative to \textit{Kepler}. We identify a subset of systems most likely to benefit from complementary dynamical constraints through \textit{PLATO} observations.

\end{abstract}

\begin{keywords}
space vehicles -- planets and satellites: detection
\end{keywords}



\section{Introduction}

Following in the footsteps of \textit{Kepler} \citep{kepler}, PLAnetary Transits and Oscillations of stars (\textit{PLATO}; \citealt{rauer_plato_2025}) will use high-precision photometry to detect and measure the transits of thousands of exoplanets. A fraction of these planets will exhibit non-linear ephemerides due to gravitational interactions with other bodies in their systems \citep{agol_detecting_2005, holman_use_2005}. These deviations from a linear ephemeris, known as transit timing variations (TTVs), most commonly arise from planet--planet interactions, but can in principle also be induced by additional companions such as exomoons \citep{kipping_transit_2009}. TTVs can therefore be used to detect non-transiting companions and to place dynamical constraints on planetary systems (e.g. \citealt{Lithwick_2012, kepler-ttvs}).

According to the NASA Exoplanet Archive \citep{christiansen_nasa_2025}, as of February 2026, 379 of the 477 planets with detected TTVs were discovered by \textit{Kepler} in its nominal mission. This follows from several aspects of the \textit{Kepler} design, including its long-duration stare of a single field, high photometric precision, and short-cadence observations of many known planet-hosting stars. Many transits are typically required to detect measurable deviations from a linear ephemeris \citep{Kane_2019}, and more still to sample a substantial portion of a TTV super-period -- the long-timescale ephemeris modulation defined by \citet{Lithwick_2012}, whose duration depends on the proximity of a planetary pair to resonance. Sampling the super-period is important for constraining the properties of perturbing companions, and \citet{agol_transit_2025} note that two well-sampled super-periods are preferable for a robust analysis. Although not all periodic TTV signals arise from near-resonant super-periods, \citet{Kane_2019} showed that strongly periodic TTV signals from confirmed \textit{Kepler} planets commonly span hundreds to more than 1000\,days.

The four-year observing baseline of \textit{Kepler}'s primary mission enabled the detection of TTVs on a range of timescales for hundreds of planets. In contrast, the Transiting Exoplanet Survey Satellite (\textit{TESS}; \citealt{tess}), with its 27-day observing sectors, has had more limited success with this method. Only 60 planets discovered by \textit{TESS} have been identified as showing TTVs as of February 2026, according to the NASA Exoplanet Archive.

The precision with which transit timings can be measured depends on both the photometric transit signal-to-noise ratio (SNR) and the observing cadence. \citet{Price_2014} showed that the uncertainty in mid-transit times is several times larger for light curves with 30-minute integrations compared to those with instantaneous sampling, and that this uncertainty scales approximately inversely with total transit signal-to-noise ratio (SNR).

The possibility of detecting TTVs with \textit{PLATO} is motivated by its expected multi-year stare durations. One current proposed observing strategy includes two two-year stares, one in each ecliptic hemisphere, although longer stares remain under consideration \citep{rauer_plato_2025}. \citet{Kane_2019} estimate that around 30 per cent of strong TTV signals from \textit{Kepler} remain detectable with just one year of data, suggesting that the \textit{PLATO} baseline will be sufficient to recover TTVs in favourable cases.

\textit{PLATO} will produce 25-second cadence imagettes for its $\sim$7,500 bright P1 sample targets in each field (see \citealt{rauer_plato_2025} for an overview of the \textit{PLATO} samples), while most of the fainter P5 sample will be observed at ten-minute cadence. Previously known exoplanet hosts may also be allocated faster cadence data products. The effect of cadence on mid-transit timing precision is therefore a relevant consideration. In addition, \textit{PLATO}'s photometric precision will vary across its target list with the number of observing cameras and due to factors such as aperture contamination and intrinsic stellar properties.

In this study, we use PlatoSim \citep{jannsen_platosim_2024} to simulate two years of time-series photometry of 152 \textit{Kepler} host stars containing 361 transiting planets, where at least one planet per system has previously detected TTVs. In addition to the instrument and detector modelling provided by PlatoSim, the end-to-end simulations incorporate simplified prototype implementations of the onboard and on-ground data processing pipelines. These prototype pipeline components, developed at LIRA Paris Observatory, approximate the expected photometric extraction and processing steps, although they do not represent the final mission pipelines.

The goal of this work is to assess whether \textit{PLATO} can recover TTV signals from these systems. The results identify which \textit{Kepler} systems are likely to benefit from additional photometric coverage and provide a benchmark for \textit{PLATO}'s expected performance in this context. 

In Section~\ref{sec:simulations}, we overview the sample properties and simulation inputs. In Section~\ref{sec:photometric_performance}, we discuss the noise characteristics of the extracted light curves. In Section~\ref{sec:fitting}, we detail the transit fitting process and introduce a timing constraint classification scheme. In Section~\ref{sec:perf_across_snr}, we discuss the timing performance across the sample and highlight systems likely to benefit from \textit{PLATO} follow-up.

\section{Simulations}
\label{sec:simulations}

\subsection{Target selection}

\begin{figure}
    \begin{center}
    \includegraphics[width=85mm]{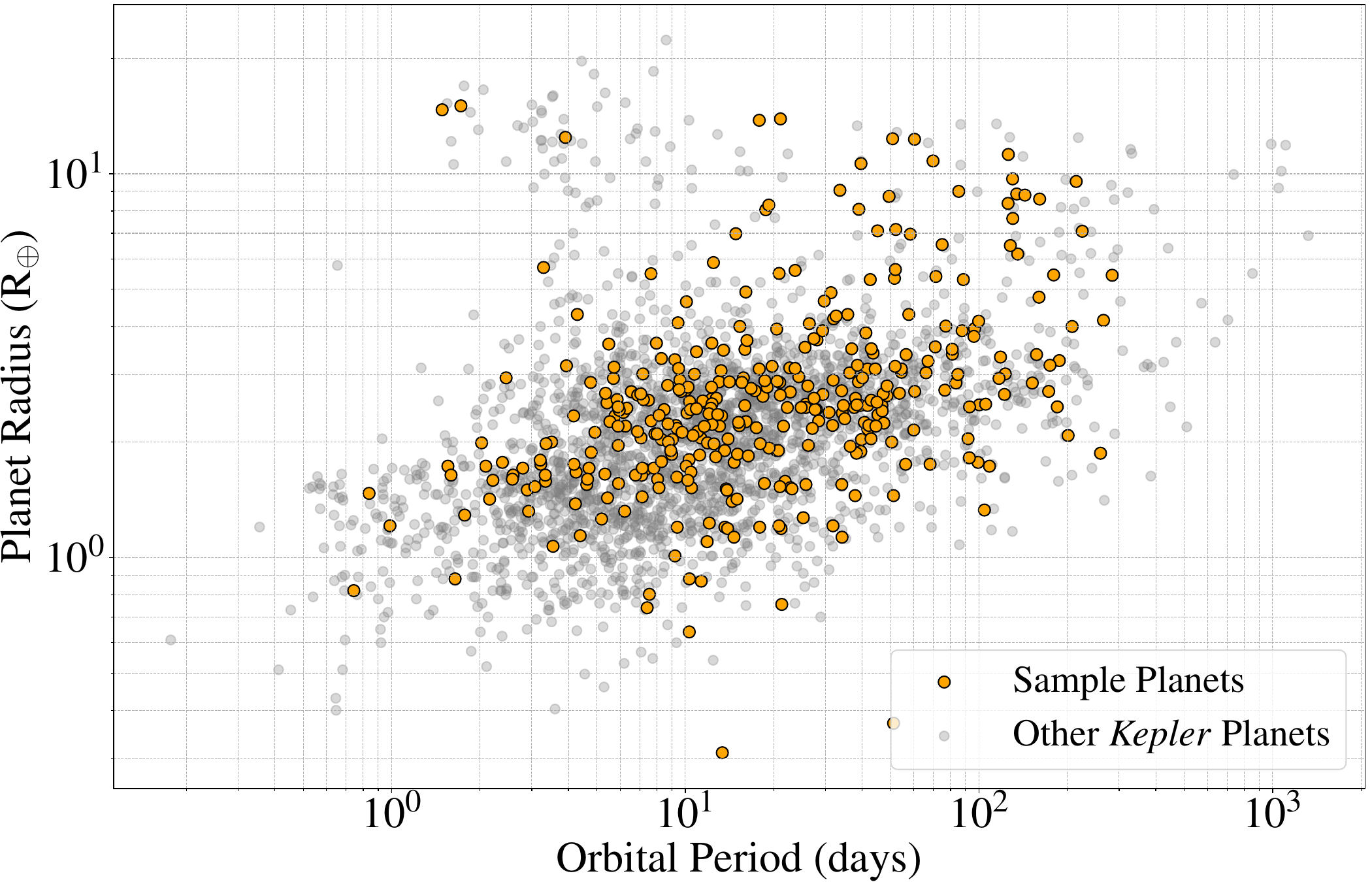}
    \end{center}
    \caption{Period-radius diagram of the 361 sample planets and the remaining 2,331 transiting exoplanets discovered by the \textit{Kepler} satellite. The data were taken from the NASA Exoplanet Archive.}
    \label{fig:period_radius}
\end{figure}

We first used the NASA Exoplanet Archive to identify all of the systems hosting at least one planet exhibiting transit timing variations and at least one planet discovered by the prime \textit{Kepler} mission. This identified 178 potential hosts. For simplicity in simulation, we limited the sample to single-star hosts, reducing the number of targets to 165. Finally, we used the host stars' temperatures and radii to exclude likely sub-giant hosts, as our stellar variability simulator cannot handle these stars. For this, we used the cutoff described in \citet{field_selection}, which is based on the work of \citet{pecaut_intrinsic_2013}. This cut reduced the number of hosts to a final sample of 152 stars, which, between them, host 361 known transiting exoplanets. We include all of these 361 planets in our sample regardless of whether they show transit timing variations. We show the sample planets' periods and radii in the context of all \textit{Kepler} discoveries in Fig. \ref{fig:period_radius}. The host brightnesses span the interval $9.77 \le V \le 16.69$~mag, with a median brightness of $V=14.67$~mag. This is marginally brighter than the median brightness of all \textit{Kepler} hosts at $V=14.83$~mag.

We note that \textit{PLATO}'s primary P1 and P5 samples include only stars brighter than $V=11$ and $V=13$~mag, respectively. Within our sample, only four stars are brighter than $V=11$~mag, and twenty are in the range $11 < V \le 13$~mag. Additionally, \citet{rauer_plato_2025} lists the photometric range of \textit{PLATO} to be $4 \le V \le 16$~mag, while 11 of our targets are dimmer than $V=16$~mag. Hence, we naturally expect the majority of hosts in our sample to be challenging targets for \textit{PLATO}.

With the sample chosen, we used a modified version of the \textsc{picsim} module in PlatoSim to generate target and contaminant catalogues for the targets. This method queries Simbad \citep{wenger_simbad_2000} to identify the named target star and find its coordinates, and then \textit{Gaia} DR3 \citep{gaiadr3} to find the target's DR3 photometric properties as well as those of contaminating sources in its vicinity. The \textit{Gaia} DR3 \textit{G}, $\textit{G}_\text{BP}$, and $\textit{G}_\text{RP}$ magnitudes are used to estimate the \textit{PLATO} passband \textit{P} magnitudes using Eq.~9 of \citet{marchiori_-flight_2019}. For each target, we opted to include all contaminants brighter than 1 per cent of the target brightness ($\Delta \text{mag} <5$) within 45~arcsec (3 pixels) of the target. This choice struck a balance between including the effect of contamination while avoiding an excessive increase in photometry extraction time, which is strongly dependent on the number of contaminants.

\begin{figure}
    \begin{center}
    \includegraphics[width=75mm]{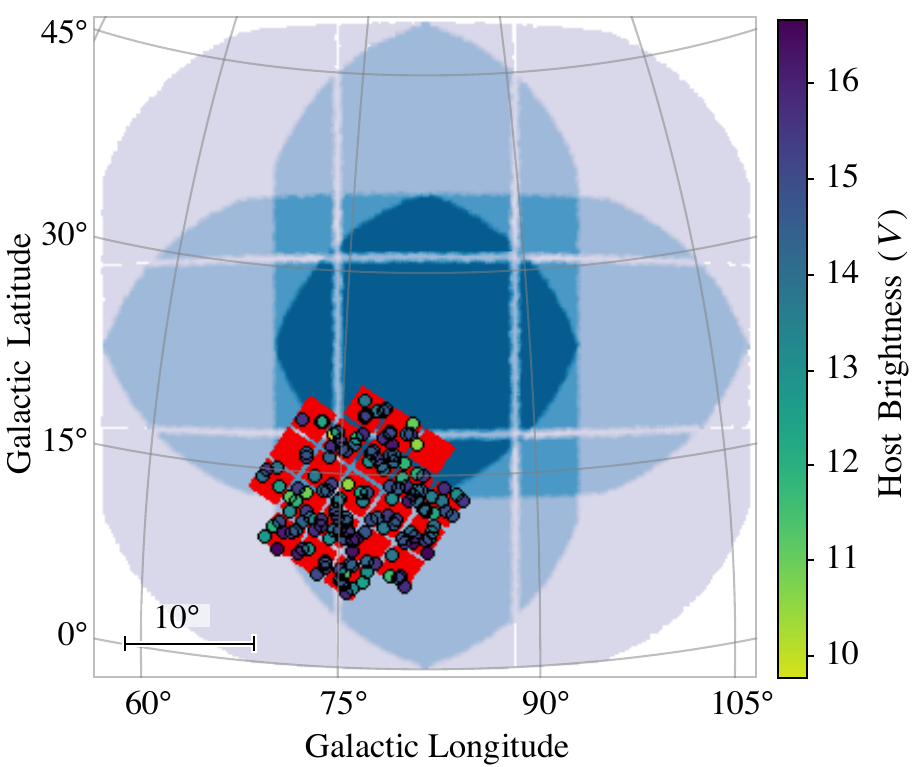}
    \end{center}
    \caption{The locations of the sample targets within the LOPN1 \textit{PLATO} and \textit{Kepler} fields. The provisional \textit{PLATO} LOPN1 field is shown in blue, with the darkness of the shaded regions indicating the number of co-pointing cameras, ranging from 6 in the lighter outer regions to 24 in the darker centre. The long-term pointing field of \textit{Kepler} is shown in red. The positions of the 152 sample stars are shown as circles.}
    \label{fig:visibility}
\end{figure}

The prime \textit{Kepler} field is encompassed in its entirety by the provisional northern \textit{PLATO} field, LOPN1 (\citealt{field_selection}, shown with \textit{Kepler} field and the sample stars in Fig.~\ref{fig:visibility}), and spans the full range of the number of co-observing, co-pointing camera groups, from one to four. Assuming LOPN1 pointing, the number of host stars in our sample observed by each possible number of normal cameras (N-CAMs) is: 6 cams -- 11, 12 cams -- 85, 18 cams -- 23, and 24 cams -- 33. For all our simulations, we assume the LOPN1 pointing, with the understanding that the final pointing may change. We note that any change is likely to be small, given the constraints on the minimum ecliptic latitude and the need to be close enough to the galactic plane to cover the required number of sample targets. Thus, it is unlikely that any proposed change would result in missing the \textit{Kepler} field entirely.

We chose to simulate all 152 systems to achieve a statistically powerful sample of TTV planet hosts, despite the considerable computational cost (discussed in Section \ref{sec:running_simulations}). We note that the similarity between the sample host and planet properties and those of the overall \textit{Kepler} population additionally permits meaningful inferences of the performance of \textit{PLATO} in following up \textit{Kepler} targets generally.

\subsection{Stellar variability}
\label{sec:stellar_variability}
We used the \textsc{varsim} module in PlatoSim to generate a realistic intrinsic stellar variability signal for each star. This module takes as inputs the stellar mass, radius, temperature, surface gravity, and [Fe/H] metallicity and is only valid for main-sequence dwarfs. Where stellar parameters were not available on the NASA Exoplanet Archive, we referred to the TICv8.2 catalogue \citep{paegert_tess_2021}. This approach retains the often greater accuracy of spectroscopically derived parameters where possible. Where metallicity data were not available, we assumed the solar value. With these parameters, \textsc{varsim} generates \textit{a} realistic stellar variability signal due to granulation based on the method of \citet{corsaro_bayesian_2013}, pulsation from \citet{kallinger_connection_2014}, and the effect of appearing and disappearing star spots modulated by stellar rotation from \citet{aigrain_testing_2015}.

A notable caveat of this approach is that for any given host star, the simulated variability may or may not accurately reflect its true variability. One natural choice may be to attempt to derive the stellar variability from the \textit{Kepler} data; however, this approach would have its own issues. For example, the problem of separating intrinsic stellar variability from instrumental noise, and that most stars will be in a different part of their magnetic cycles by the time they are observed by \textit{PLATO} compared to the point at which they were observed by \textit{Kepler}. Additionally, rather than attempting to generate some modelled intrinsic variability for all the contaminating stars, we modelled them as constant sources. We expect PSF-fitting photometry to mitigate much of the impact of varying contaminants in real data, although targets with aperture photometry will likely be more severely affected.

This method is a practical approach for simulating large numbers of stars, which, in addition to enabling statistical inferences, also enables meaningful predictions for individual systems.

\subsection{Transit injection}

We again used \textsc{varsim} and the NASA Exoplanet Archive to simulate the noise-free transit light curves for each star. The transit modeller within \textsc{varsim} takes as inputs the ephemeris time, orbital period, inclination, eccentricity, argument of periastron, planet mass, and planet radius in addition to the stellar parameters listed in Section \ref{sec:stellar_variability}. It implements \textsc{batman} \citep{kreidberg_batman_2015} for primary transits, \textsc{spiderman} \citep{louden_spiderman_2018} for secondary occultations, and \textsc{ldtk} \citep{parviainen_ldtk_2015} for generating \textit{PLATO} bandpass-specific limb darkening coefficients. Doppler beaming and ellipsoidal variation are implemented in \textsc{pyastronomy} \citep{pya}, but since the masses of \textit{Kepler} planets are usually unknown, we assumed zero mass for all planets, thereby removing the impact of these (already minor) effects.

We adopted planetary periods, inclinations, eccentricities, and radii from the default parameter sets provided by the NASA Exoplanet Archive. In cases where the argument of periastron was not listed -- typically due to the difficulty of measuring this parameter at low eccentricity -- we assumed a constant value of $90^\circ$. When eccentricities or inclinations were absent, values of zero and $90^\circ$ were assumed, respectively. For Kepler-103~c, the default stellar and planetary parameters listed in the archive, drawn from \citet{bonomo_cold_2023}, yield a transit impact parameter greater than 3, implying a non-transiting geometry. The inclination reported by \citet{bonomo_cold_2023}, $87.704^{+0.12}_{-0.055}$$^\circ$, appears to be a transcription error from the source values reported by \citet{dubber_using_2019}, differing by exactly 2$^\circ$. This inconsistency was identified only after the simulations and subsequent analysis had been completed, and Kepler-103~c was therefore excluded from the final sample, reducing the total to 360 planets.

Since it is not currently known exactly when \textit{PLATO} will begin observations of its Northern field (if at all), and a large fraction of our sample planets don't have ephemeris times listed on the NASA Exoplanet Archive, we uniformly sampled random ephemeris times between zero and each planet's orbital period. 

We combined the (potentially multiple) noiseless transit light curves for each star with their stellar variability light curves by multiplying them together to produce two years of intrinsic photometric variability for each system.

\subsection{Payload simulation}

The final ingredient needed to generate CCD imagettes is the \textsc{payload} module, which manages the telescope systematics. This module injects alignment errors between the cameras and the optical bench, pointing repeatability errors for each observing quarter, thermoelastic distortion of the optics, a realistic distribution of observational downtime, and gain-thermal transients that increase the CCD temperature after quarterly rolls. We used all of the default parameters of this module (see \citealt{jannsen_platosim_2024}) to generate eight quarters of payload systematics.

\subsection{Photometry extraction and simulation runs}
\label{sec:running_simulations}

We used the built-in \textsc{python} wrapper of PlatoSim, \textsc{platonium}, to run the simulations. This combines the outputs of the \textsc{picsim}, \textsc{varsim}, and \textsc{payload} modules, runs the simulation for one star over one quarter (of a year) as observed by one camera, and then applies the chosen photometry pipeline.

We do not yet know precisely which data product will be produced for these targets; therefore, we adopted a mixed approach to photometry extraction. Firstly, we simulated the use of the 25-second cadence imagettes for on-ground PSF-fitting photometry via the L1 photometry pipeline. This is the treatment that the P1, P2, P4 and $\sim$3.7 per cent of the P5 sample targets will receive. Secondly, on the same imagettes, we simulated onboard aperture photometry using the method of \citet{marchiori_-flight_2019}, which constructs photometric precision-maximizing apertures composed of whole pixels that are updated periodically throughout the quarter. This method is analogous to the expected treatment of the majority of the P5 sample stars for which bandwidth limitations preclude imagette downlinking.

In both cases, the treatment of the PSF is important. For stars subject to PSF fitting, point spread functions with subpixel resolution are obtained via a microscanning approach (see \citealt{samadi_plato_2019} for details). This method involves acquiring a series of imagettes of the target that are shifted by subpixel displacements following an Archimedean spiral, and combining these data using a regularized least-squares approach with a positivity constraint on the solution. In our simulations, \textsc{platonium} orchestrates the spiral observations within PlatoSim at the beginning of each observing quarter, while the L1 proto-pipeline performs the PSF inversion and fitting.

For stars for which only aperture photometry is performed, knowledge of the PSF is still required -- in order to define the photometric apertures. In the nominal processing, PSFs are reconstructed for reference stars across the field of view and interpolated to estimate the PSF at a given focal-plane position. In our simulations, however, we do not perform this reconstruction step; instead, we use the theoretical PSFs provided by PlatoSim.

We initially ran 29,184 simulations -- 152 stars observed by 24 cameras over 8 quarters. We ran simulations from mission quarters 9 to 16 in \textsc{platonium} to account for the expected instrumental degradation between the beginning of the mission and when the Northern field is observed. Since we did not know in advance which or how many cameras the targets would be visible to, we ran the simulations for all 24 cameras. Simulations for which the target was outside the field of view terminated early. These simulations took approximately seven weeks spread across three systems, each equipped with two 32-thread Intel Xeon Silver 4514Y processors, for a total of around 200,000 CPU hours. The execution time was dominated by the computational cost of the L1 PSF-fitting photometry proto-pipeline.

Additionally, we ran an identical set of simulations using only aperture photometry, with all contaminants removed, to quantify the mean photometric contamination level per simulation and per mask update. These simulations took around three days using one of the above-described machines for a total of around 4,000 CPU hours. For each mask update, we subtract the mean contamination level from the contaminant-present aperture photometry. This produces light curves corrected for contamination in the mean signal, while retaining the noise impact of photometric contaminants. In the real data, contamination will be addressed via the spacecraft jitter correction, which re-scales the measured flux to the level expected with only the target. These corrections can be validated by measuring the dilution of exoplanet transits relative to the catalogued transit depths. We did not utilize the jitter or long-term drift corrections in the present work.

The PSF-fitting photometry failed for one eighth of the targets -- 19 stars hosting 50 planets. These 19 stars all have $V\ge14.79$~mag, and the mean and standard deviation of the failed PSF-fitting host brightnesses in \textit{V} are 15.79 and 0.53~mag, respectively. Failed extraction in the current pipeline implementation appears to be associated with a combination of low target brightness and high levels of photometric contamination. We note, however, that PSF-fitting photometry is generally expected to outperform aperture photometry in crowded fields and for faint sources. This has been demonstrated for \textit{Kepler} -- which used similarly undersampled PSFs -- by \citet{libralato_psf-based_2016, libralato_psf-based2_2016}. We therefore expect comparable improvements to be achievable. The PSF-fitting method used in the present work was primarily developed for and tested against stars with $V\le11$~mag, which may contribute to the reduced performance observed for dimmer targets. However, given that the faint ($V\le16$~mag) P4 M-dwarf sample stars will receive the same 25-second imagette data product as the bright P1/P2 samples, we expect the final implementation of the PSF-fitting routine to more effectively handle dim, crowded targets. Despite the limitations of the current PSF-fitting photometry implementation, aperture photometry was successfully extracted for the entire sample.

Each of the four CCDs in a \textit{PLATO} camera is read out with a 6.25-second offset, allowing an effective cadence finer than 25 seconds. In principle, this enables a combined light curve with 6.25-second resolution when data from all four camera groups are used. However, only transits with very high signal-to-noise ratios would significantly benefit from such high temporal resolution, which was not the case for the overwhelming majority of our targets. Additionally, each camera group produces an independent light curve that would require separate detrending before combination, with the risk of introducing additional noise. Thus, we chose to average the light curves from the four groups to produce a final light curve with a uniform 25-second cadence, which remains superior to the shortest cadence of the \textit{Kepler} data (58.85 seconds).

\subsection{Detrending}
\label{sec:detrending}

We chose to use the biweight detrending method implemented in \textsc{wotan} \citep{hippke_wotan_2019} with a window length of one day based on its good performance in \citet{canocchi_discovering_2023}. Since we knew from the simulation inputs the transit times and durations, we masked out all of the transits before detrending. This work focuses on the measurement and detectability of known transits, making this approach readily applicable to real data. For each star, we applied this method to both the aperture photometry and, where available, the PSF-fitting photometry 25-second cadence light curves.

\section{Photometric performance}
\label{sec:photometric_performance}

We measured the standard deviations of the 3$\sigma$-clipped out-of-transit portions of the 25-second cadence detrended light curves derived from both photometry methods. We then binned the data to a 1-hour cadence by taking the mean average of every consecutive $3600/25=144$ points, weighted by their inverse variances and additionally measured the scatter in the binned light curves. 

\subsection{Comparison of PSF and aperture photometry}

\begin{figure}
    \begin{center}
    \includegraphics[width=85mm]{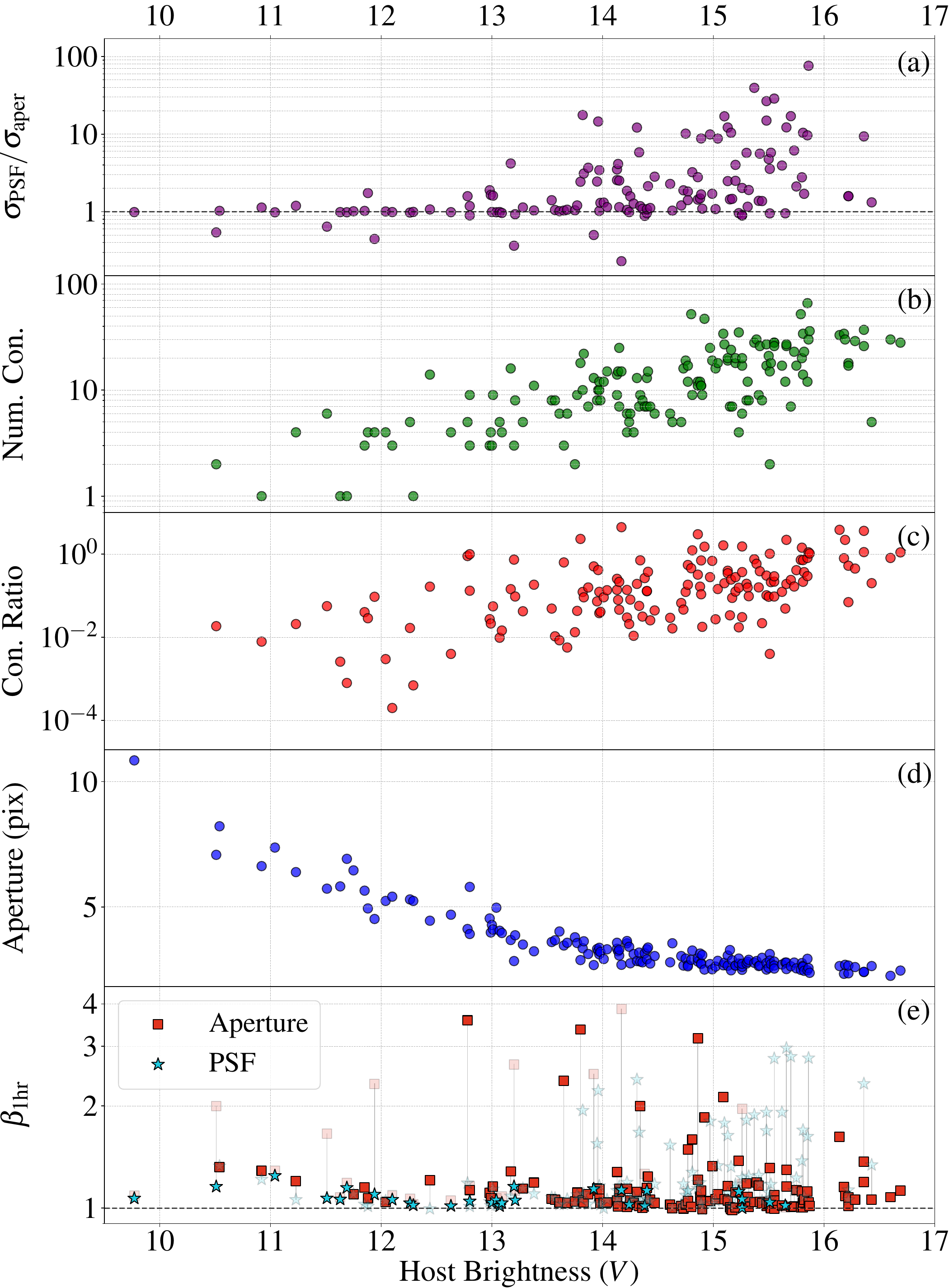}
    \end{center}
    \caption{Plots of, as a function of host brightness: \textbf{(a)} The ratio of the PSF-photometry-derived to aperture-photometry-derived 1-hour photometric noise. \textbf{(b)} The number of contaminating stars within 45\,arcsec brighter than 5 magnitudes dimmer than the targets. \textbf{(c)} The mean ratio of contaminant to target flux within the photometric apertures. \textbf{(d)} The mean size in pixels of the \protect\cite{marchiori_-flight_2019} derived photometric apertures. \textbf{(e)} The noise redness factor $\beta_\mathrm{1hr}$ as defined in Eq. \ref{eq:beta} for each photometry method. For each star, the value associated with the photometry method that yields the lower 1-hour noise is shown with greater opacity.}
    \label{fig:contamination}
\end{figure}

Substantial differences were observed between the noise-to-signal ratios (NSRs) obtained from PSF-fitting and aperture photometry. Across our sample, the ratio $\sigma_{\mathrm{PSF}}/\sigma_{\mathrm{aper}}$ spans more than two orders of magnitude, ranging from $0.23$ (PSF lower noise) to $76$ (aperture lower noise), as shown in panel (a) of Fig.~\ref{fig:contamination}.

The relative performance of the two methods varies from star to star and reflects a combination of target brightness and local crowding (see panels (b) and (c) of Fig.~\ref{fig:contamination} showing the variation of the number of contaminants and the relative amounts of contaminating flux with host brightness). Because the L1 PSF-fitting pipeline is under active development, we do not attempt to interpret these differences in detail, but we note that relative performance is most strongly correlated with the number of contaminants, rather than the contamination ratio. For each target, we adopt the photometry method that yields the lower 1-hour noise. This ensures that our timing forecasts reflect the best-performing extraction method available for each star within the current simulation framework. In total, 28 stars achieve better precision with PSF-fitting photometry and the remaining 124 stars achieve better precision with aperture photometry.

\subsection{Comparison with modelled and \textit{Kepler} noise}
\label{sec:model_noise_comparison}

\begin{figure}
    \begin{center}
    \includegraphics[width=85mm]{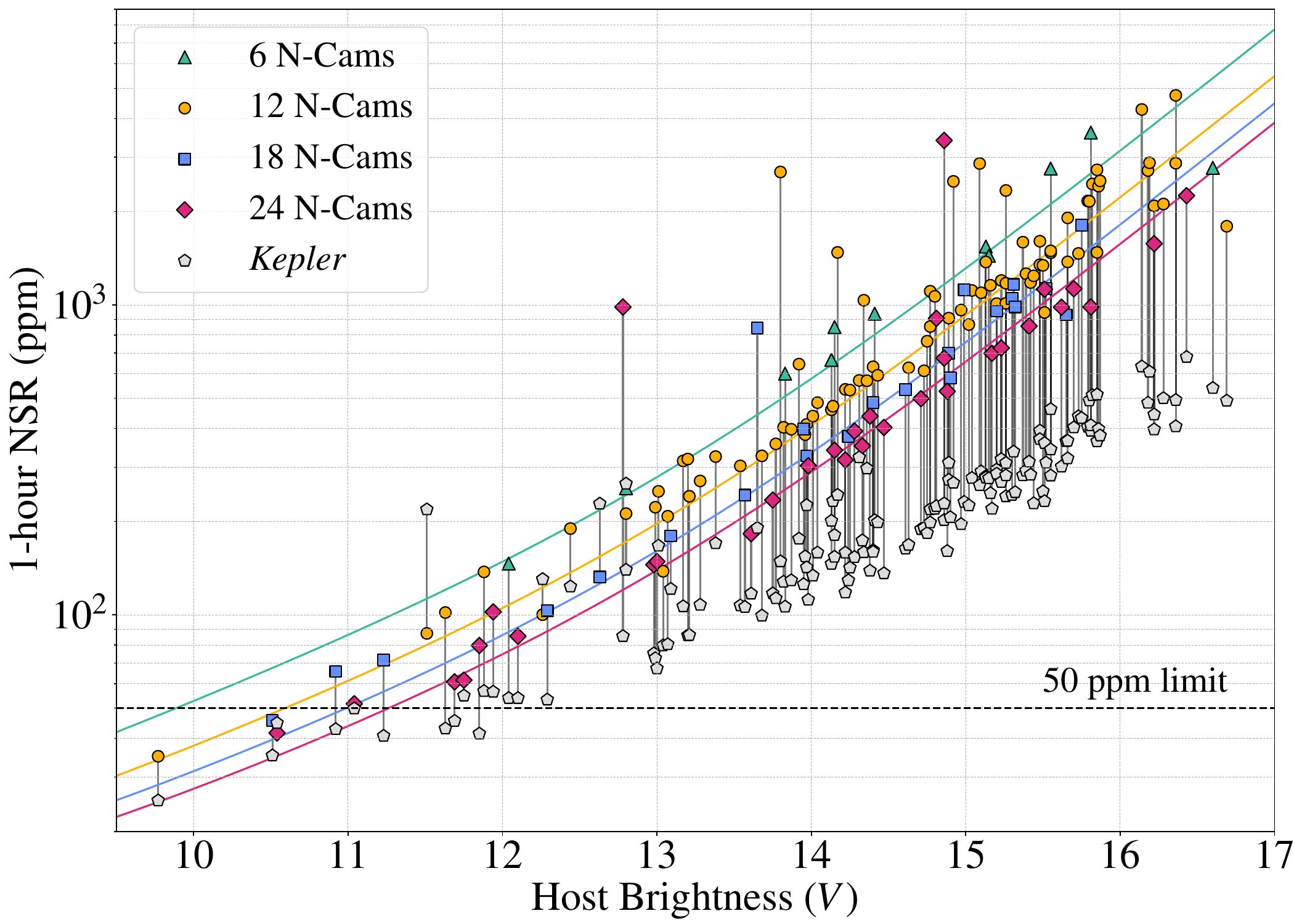}
    \end{center}
    \caption{Noise-magnitude plot for the sample of 152 stars. The noise is the measured 3$\sigma$-clipped scatter in the detrended and then 1-hour binned out-of-transit light curves. Of the two noise-to-signal ratios calculated from the light curves of the two photometry extraction methods, the lowest noise result is shown for each star. The stars observed by 6, 12, 18, and 24 cameras are denoted by triangles, circles, squares, and diamonds, respectively. The scatter in the real \textit{Kepler} data is calculated in the same way, shown as pentagons with lines connecting the real \textit{Kepler} and simulated \textit{PLATO} scatters for the same stars. The modelled \protect\citep{cabrera_assessment_2026} \textit{PLATO} performances for 6, 12, 18, and 24 cameras are shown as solid lines. The 50\,ppm threshold for Earth-analogue detection is indicated by a horizontal dashed line.}
    \label{fig:brightness_noise}
\end{figure}

\begin{figure}
    \begin{center}
    \includegraphics[width=85mm]{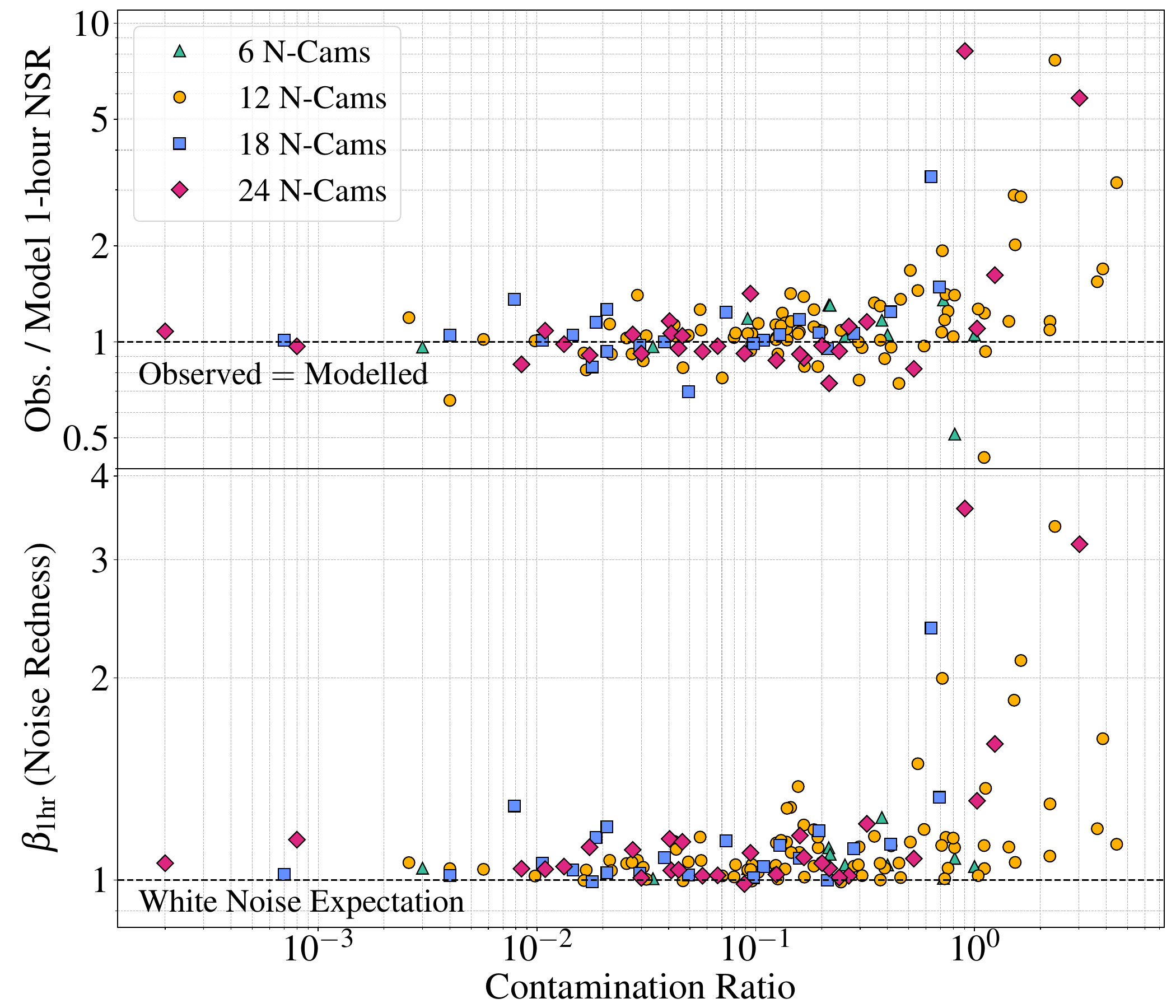}
    \end{center}
    \caption{\textbf{Top:} Ratio of the measured 1-hour noise-to-signal ratios to those predicted by the model of \protect\citet{cabrera_assessment_2026}, shown as a function of the contamination ratio. The stars observed by 6, 12, 18, and 24 cameras are denoted by triangles, circles, squares, and diamonds, respectively. \textbf{Bottom:} Noise redness, $\beta_\mathrm{1hr}$, as defined in Eq.~\ref{eq:beta}, shown as a function of the contamination ratio (the mean ratio of contaminant to target flux in the photometric apertures).}
    \label{fig:noise_excess_contamination}
\end{figure}

In Fig. \ref{fig:brightness_noise}, we show the 1-hour noise-to-signal ratio (NSR) as a function of host brightness and camera number, along with the beginning-of-life (BOL) condition `Quick noise model' of \citet{cabrera_assessment_2026}\footnote{The noise model of \citet{cabrera_assessment_2026} is parameterized in \textit{PLATO} \textit{P} magnitudes. To maintain consistency with the use of \textit{V} magnitudes throughout this work, we adopt the \textit{V} magnitude to electron-flux conversion given in Appendix~A, Eq.~1 of \citet{rauer_plato_2025}.}. Additionally, the ratio of observed to modelled noise is shown against contamination ratio in the upper panel of Fig.~\ref{fig:noise_excess_contamination}.

Across our sample, we observe reasonable agreement with the model predictions across the range of host brightnesses. The median ratio of observed-to-predicted noise is 1.061, indicating that the model marginally underpredicts noise levels on average. This noise excess is not generally due to stellar variability. By applying the same detrending approach as in Section~\ref{sec:detrending} to the raw, injected stellar signals, we find that the levels of intrinsic stellar variability on one-hour timescales fall in the interval $6.2\,\mathrm{ppm}\le \sigma_{\star,\mathrm{1hr}}\le32.4\,\mathrm{ppm}$. Subtracting in quadrature the stellar noise from the total one-hour noise for each star reduces the median noise ratio to 1.059. Thus, the stellar component makes a negligible contribution to the noise of all but the brightest stars in our sample. As shown by \citet{cabrera_assessment_2026}, the noise properties of \textit{PLATO} are expected to degrade between its BOL and end-of-life (EOL) states. This evolution is modelled in PlatoSim and may explain the remaining noise excess.

Two notable outliers are the dimmest stars in \textit{V}, which exhibit substantially lower noise than predicted by the model. These systems, Kepler-1624 and Kepler-267, also have the reddest $\textit{V}-\textit{P}$ colours in the sample, with values of 1.40 and 1.66~mag respectively, compared to the sample median of 0.44~mag. Their large $\textit{V}-\textit{P}$ colours imply that they are significantly brighter in the \textit{PLATO} bandpass than their \textit{V} magnitudes would suggest, naturally leading to lower photometric noise.

Conversely, a notable fraction of stars show significantly larger noise levels than predicted. This effect appears to be correlated with the flux contamination ratios (shown in the upper panel of Fig. \ref{fig:noise_excess_contamination}) with a turn-on at ratios $\gtrsim0.5$.

We note that the quick noise model of \citet{cabrera_assessment_2026} assumes a fixed photometric aperture size. However, as shown in panel (d) of Fig.~\ref{fig:contamination}, the photometric-precision-optimizing aperture selection method of \citet{marchiori_-flight_2019} naturally selects smaller apertures for dimmer stars. One might therefore expect the model to overestimate the noise levels of faint targets, since smaller apertures include smaller contributions from sky background and read noise. In practice, however, this effect is not clearly observed, likely because any reduction in read noise and sky-background noise is offset by lower integrated flux and the generally higher contamination levels associated with fainter stars.

\begin{figure}
    \begin{center}
    \includegraphics[width=85mm]{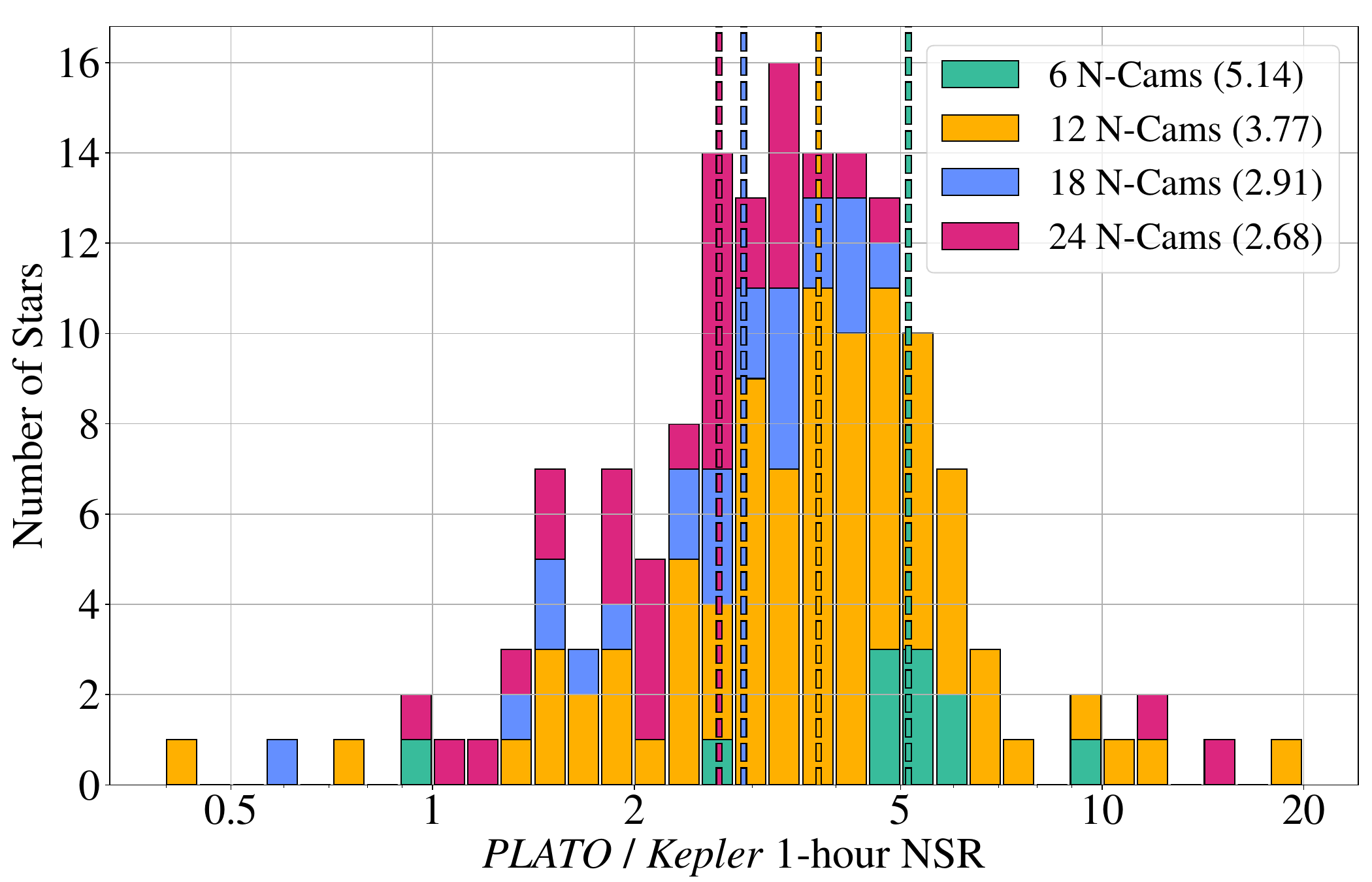}
    \end{center}
    \caption{Distribution of the relative noise performance of \textit{PLATO} and \textit{Kepler} for the sample stars by number of co-pointing cameras. Vertical dashed lines indicate the median noise ratio for each number of cameras (numerically shown in the legend).}
    \label{fig:plato_vs_kepler}
\end{figure}

Fig. \ref{fig:plato_vs_kepler} shows the distribution of the relative noise performance of \textit{PLATO} and \textit{Kepler} for the sample of host stars. The \textit{Kepler} noise levels were derived in largely the same manner as for the \textit{PLATO} light curves: by detrending, where possible, the entire available short-cadence data and measuring the 3$\sigma$-clipped 1-hour scatter. Long-cadence data were used when short-cadence data were unavailable. The median relative noise performance scales as expected with the inverse square root of the number of cameras, ranging from 2.68 times the \textit{Kepler} noise for 24 cameras to 5.14 times the noise for six cameras.

\subsection{Correlated noise}

Under the assumption of white noise, binning by a factor $n$, or mean-averaging every consecutive $n$ data points, lowers the root-mean-square (RMS) scatter by a factor $\sqrt{n}$. As defined by \citet{winn_transit_2007}, we take as a measure of correlation in the noise the ratio of the measured noise in the 1-hour binned out-of-transit light curve to the level expected by reducing the 25-second noise by a factor $\sqrt{n}=12$. Formulaically,

\begin{equation}
\label{eq:beta}
\beta_\mathrm{1hr}=\frac{12\sigma_\mathrm{1hr}}{\sigma_\mathrm{25s}},
\end{equation}

where a value $\beta_\mathrm{1hr}=1$ is consistent with the expectation for white-noise. A plot of $\beta_\mathrm{1hr}$ against host brightness for each star and for each photometry method is shown in panel (e) of Fig.~\ref{fig:contamination}. 

The median $\beta_\mathrm{1hr}$ across the sample is 1.063 and the 90th percentile value is 1.297, suggesting that most stars are dominated by white noise up to 1-hr timescales, which implies that the out-of-transit variability is slow enough to be captured by the 1-day window detrending. 

There are some significant outliers (nine stars with $\beta_\mathrm{1hr}>1.5$). We find that the best predictor of non-white noise (a large $\beta_\mathrm{1hr}$) is simply the flux contamination ratio. This is illustrated in the lower panel of Fig. \ref{fig:noise_excess_contamination}. Similar to the noise excess, significant red noise appears to occur at contamination ratios $\gtrsim0.5$, although substantial variation in $\beta_\mathrm{1hr}$ remains even at large contamination ratios.

Additionally, we investigated the impact of stellar variability on the measured noise redness. Using the detrended, injected intrinsic variability described in Section~\ref{sec:model_noise_comparison}, we calculated a \textit{stellar-only} $\beta_\mathrm{1hr}$ value for each star. The resulting 16th, 50th, and 84th percentiles over the stellar sample are $[3.30, 3.64, 4.39]$, indicating that the intrinsic stellar variability is strongly red. However, for this redness to measurably affect the extracted photometry, the stellar contribution must also constitute a significant fraction of the total noise budget.

To estimate the contribution from the instrument alone, we derived an \textit{instrumental-only} $\beta_\mathrm{1hr}$ for each star by subtracting the stellar noise in quadrature from the total noise at both the 25-second and 1-hour cadences. We find that the instrumental-only $\beta_\mathrm{1hr}$ is only significantly closer to unity than the total $\beta_\mathrm{1hr}$ -- which we define as $\beta_\mathrm{1hr}-1$ decreasing by at least a factor of two -- for four stars, all brighter than $V=12$~mag. The median instrumental-only redness is 1.059, only marginally smaller than the median value for the total noise. This suggests that, although stellar variability is intrinsically highly red, it contributes only weakly to the overall red noise budget for most stars in the sample. Consequently, the observed redness of the extracted photometry is dominated primarily by instrumental systematics except for the brightest targets, where the stellar contribution becomes appreciable.

\section{Transit Fitting}
\label{sec:fitting}

In the following section, we describe the procedure used to infer mid-transit times from simulated \textit{PLATO} light curves. Our aim is not to detect transits independently, but instead to quantify the precision with which individual transit epochs can be measured given strong prior knowledge of the systems from \textit{Kepler} or from combining transits from \textit{PLATO}. Accordingly, we treat the transit shape and ephemeris as known to a good approximation, and focus on the conditional inference of the mid-transit time for each observed transit.

\subsection{Light curve preparation}

For each planet, we prepared up to ten individual transit light curves for fitting, balancing statistical completeness against computational cost. Using the injected mid-transit times, we extracted segments from the detrended 25\,s cadence light curves spanning $\pm 1.5$ transit durations around each expected transit centre, for all transits occurring within the two-year observing window. This window size provides a sufficient out-of-transit baseline for normalization while avoiding the inclusion of unnecessary stellar variability. For each star, we adopted the photometric extraction method that yielded the lowest 1-hour noise metric.

Within each extracted window, we masked any transits of other known planets in the same system to avoid contamination. We then constructed a vetted sample of transit windows by requiring at least 95\,per\,cent temporal coverage of the transit itself and at least 50\,per\,cent temporal coverage of both the pre- and post-transit baseline regions. From this vetted set, we selected up to ten transits per planet, approximately uniformly distributed over the observing baseline. For planets with fewer than ten suitable transits, all vetted transits were retained. Finally, we applied a 5$\sigma$ clip to the injected-model-normalized flux to correct for the effect of cosmic rays, which is particularly impactful on aperture photometry. We note that this is a simplified, ad hoc approach to outlier detection that does not reflect the most up-to-date methods in the photometry pipelines.

\subsection{Transit model and likelihood}

For each extracted transit window, we model the observed flux using a fixed-shape limb-darkened transit light curve computed with the \textsc{batman} package, adopting a quadratic limb-darkening law. All orbital and geometric parameters are held fixed at their injected values, including the orbital period, scaled semi-major axis, scaled planet radius, inclination, eccentricity, argument of periastron, and limb-darkening coefficients. This approach isolates the inference of the mid-transit time while avoiding degeneracies with poorly constrained shape parameters.

The free parameters in the fit are the mid-transit time $t_0$, a multiplicative scaling factor applied to the transit depth $d$, a multiplicative flux baseline $B$, and an additive jitter term $s$ that accounts for potential underestimation of the nominal photometric uncertainties. Allowing for a depth scaling accommodates small mismatches between the injected and observed transit depths due to residual uncorrected contamination.

Assuming Gaussian-distributed photometric errors, the likelihood for a set of flux measurements $f_i$ at times $t_i$ is given by
\begin{equation}
\ln \mathcal{L} = -\frac{1}{2}\sum_i \left[\frac{(f_i - m_i)^2}{\sigma_i^2+s^2} + \ln\left(2\pi(\sigma_i^2+s^2)\right)\right],
\end{equation}
where $m_i \equiv m(t_i; t_0,d,B)$ is the model flux and $\sigma_i$ are the nominal per-point uncertainties. In the case of the PSF-fitting photometry, the per-point uncertainties are formally estimated by -- and a direct output of -- the L1 proto-pipeline. Conversely, uncertainties were estimated for the aperture photometry derived using the method of \citet{marchiori_-flight_2019} by calculating, for each mask update, the 3$\sigma$-clipped standard deviation of the detrended, transit-masked light curves, since formal uncertainties were not provided. The additional variance term $s^2$ is fitted simultaneously to ensure that timing uncertainties are not artificially underestimated in the presence of imperfect noise estimation.

Priors are imposed as follows. The mid-transit time $t_0$ is assigned a uniform prior within a window of $\pm\,\mathrm{duration}$ around the predicted ephemeris. This choice reflects reasonable prior knowledge obtained either through the propagation of \textit{Kepler} ephemerides or the stacking of \textit{PLATO} transits while allowing the posterior to be dominated by the likelihood function. The jitter parameter is sampled uniformly in $\ln s$ over the range $[-20,1]$. The depth scaling factor $d$ follows a Gaussian prior centred close to unity, with an exact mean and width calibrated empirically from high signal-to-noise simulated transits and separately for the PSF-fitting and aperture photometry. This strong prior on the depth reflects confidence in both the knowledge of the transit shape and the accuracy of the contamination correction, and is necessary to prevent the sampler from focusing on high-likelihood but physically implausible solutions. The baseline parameter $B$ is assigned a uniform prior over the interval $[0.9,\,1.1]$.

\subsection{Parameter estimation and posterior sampling}

Posterior distributions for the model parameters were explored using Markov Chain Monte Carlo (MCMC) sampling. For each transit window, we employed the ensemble sampler implemented in the \textsc{emcee} package \citep{foreman-mackey_emcee_2013}. The sampler was initialized with 32 walkers distributed around an initial parameter estimate corresponding to the injected mid-transit time, unit depth scaling, unit baseline, and a jitter term set by the median photometric uncertainty in the window.

Each chain was evolved for 10,000 steps, with the first 1000 steps discarded as burn-in. Convergence was assessed by verifying that the autocorrelation lengths of all parameters were at most 1/50 the retained chain length (satisfied by $\sim$99 per cent of fits).

For each parameter, posterior summaries are reported using the median and central credible intervals. In particular, the uncertainty on the mid-transit time is quantified using half the interval between the 16th and 84th percentiles of the marginal posterior distribution. We retain the full posterior samples for subsequent analysis, including the assessment of posterior structure and model support described in the following sections.

\subsection{Classification of timing constraints}

Not all transit fits yield mid-transit time estimates that are equally informative. In particular, low signal-to-noise transits may produce posterior distributions that appear well behaved despite being dominated by the imposed priors or by chance alignment with noise. To distinguish genuinely informative timing constraints from such cases, we classify each transit fit based on the structure of its posterior distribution and the degree to which the transit model is preferred to a flat, no-transit model.

Our classification scheme proceeds hierarchically, and each transit is assigned to one of four mutually exclusive classes, as described below.

\subsubsection{Prior-limited}
Fits are classified as \emph{prior-limited} when the posterior distribution of the mid-transit time $t_0$ is strongly influenced by the bounds of the uniform timing prior. Specifically, if the posterior median is consistent with the lower or upper prior boundary to within three times the lower or upper timing uncertainty, respectively, the timing constraint is considered prior-dominated and the transit is assigned to this class. This criterion captures cases where the apparent constraint arises primarily from the imposed timing window rather than from information in the data and is often associated with a mid-transit timing error (extracted minus injected time) much smaller than the calculated timing uncertainty.

\subsubsection{Low significance}
For fits that are not prior-limited, we assess whether the data provide meaningful support for a transit-shaped signal within the transit window. This is quantified using the difference in maximum log-likelihood, $\Delta \ln \mathcal{L}$, between the transit model and a null (constant-flux) model. In the null model, we fix the jitter term to the best-estimate value from the transit model fit. This prevents high-likelihood null model fits from being achieved through unrealistic inflation of per-point uncertainties, which can consume the significance of the transit model relative to the null model. Transits with $\Delta \ln \mathcal{L} < 5$ are classified as \emph{low significance}, indicating that the transit model is not strongly preferred over noise within the fitted window.

\subsubsection{Asymmetric high significance and symmetric high significance}
Transits with $\Delta \ln \mathcal{L} \ge 5$ are considered to have significant support for a transit-shaped signal. Among these, we further distinguish cases based on the symmetry of the marginal posterior distribution of $t_0$. To quantify this, we define an asymmetry metric based on posterior quantiles,
\begin{equation}
A = \frac{q_{84} + q_{16} - 2q_{50}}{q_{84} - q_{16}},
\end{equation}
where $q_{16}$, $q_{50}$, and $q_{84}$ denote the 16th, 50th, and 84th percentiles of the posterior, respectively. This quantity provides a measure of skewness in the interval $[-1,1]$.

Transits with $|A| > 0.3$ are classified as \emph{asymmetric high significance}, typically corresponding to cases where the timing constraint is strongly one-sided or multi-modal. Transits with $|A| \leq 0.3$ are classified as \emph{symmetric high significance} and represent the most robust per-transit timing measurements, exhibiting approximately unimodal and symmetric posterior distributions for $t_0$.

This classification forms the basis for the population-level analysis of transit timing performance presented in the following section.

\section{Transit timing performance and classification across SNR}
\label{sec:perf_across_snr}

In the following section, we refer to the transit signal-to-noise ratio (SNR), which we define in the canonical, albeit simplistic \citep{kipping_snr_2023}, fashion:

\begin{equation}
\mathrm{SNR}=\frac{d_\mathrm{transit}\sqrt{t_\mathrm{duration}/\mathrm{1hr}}}{\sigma_\mathrm{1hr}},
\end{equation}

where $d_\mathrm{transit}$ is the maximum transit depth, $t_\mathrm{duration}$ is the total (first to fourth contact) transit duration, and $\sigma_\mathrm{1hr}$ is the photometric noise-to-signal ratio in one hour.

\subsection{Transit SNRs of the sample planets}

\begin{figure}
    \begin{center}
    \includegraphics[width=85mm]{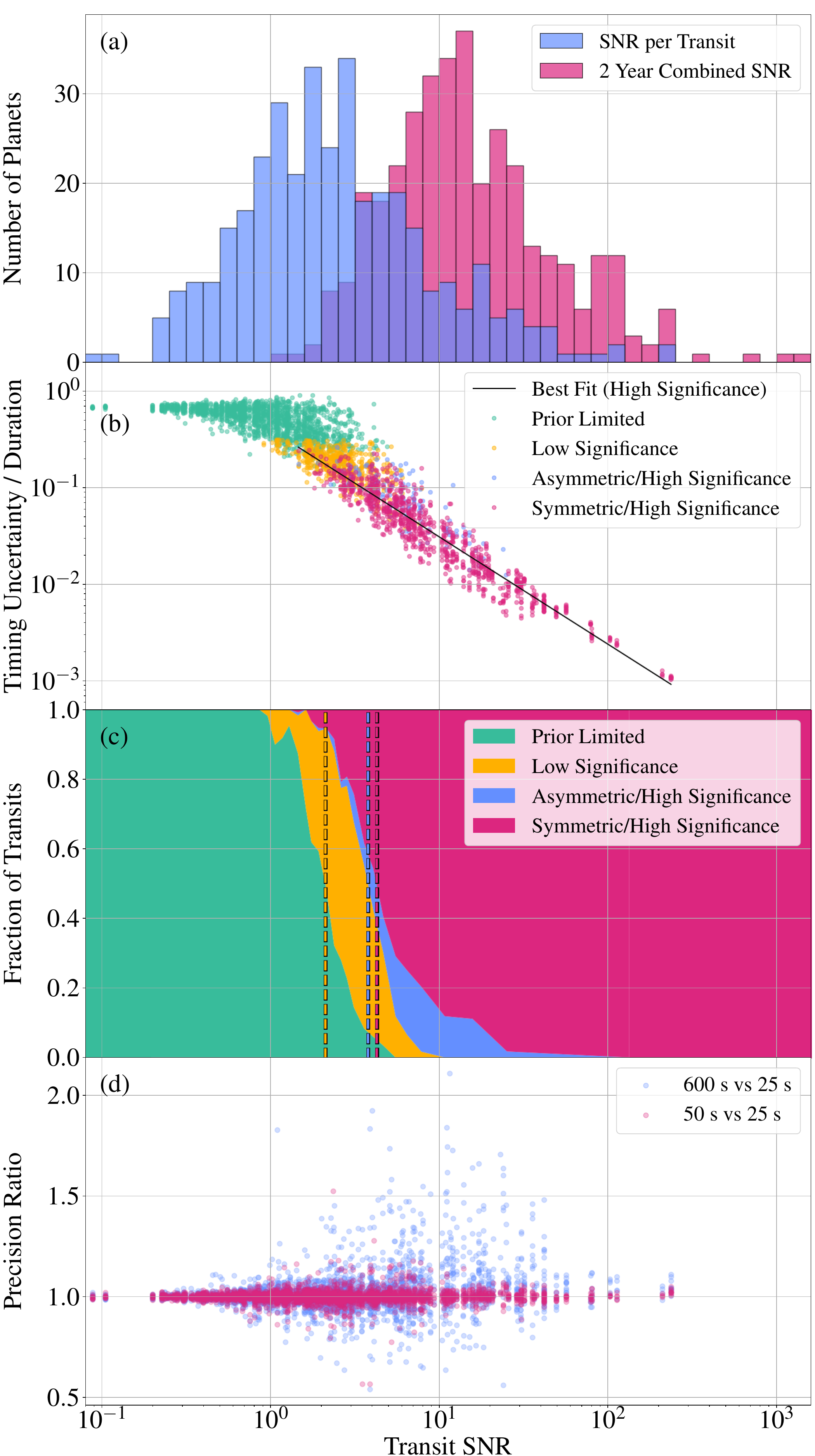}
    \end{center}
    \caption{\textbf{(a)} Histogram distribution of the expected transit signal-to-noise ratios within the planet sample, both from one single transit and also when combining all the transits expected in a two-year window. \textbf{(b)} The ratio of the mean timing uncertainty (half the 16th--84th percentile range) to the transit duration for each individual transit fit. A linear (log-log) fit to the two high-significance classes is shown as a solid line. \textbf{(c)} The (non-cumulative) fraction of transit fits falling into each class as a function of transit SNR. The SNR at which 50 per cent of transit fits achieve at least a given class are shown as vertical dashed lines. \textbf{(d)} The ratio of the timing precisions arising from 600- and 25-second sampled and 50- and 25-second sampled data.}
    \label{fig:fit_statistics}
\end{figure}

The distribution of expected SNRs across the planet sample is shown in panel (a) of Fig.~\ref{fig:fit_statistics}. In the \textit{Kepler} mission's Transiting Planet Search, folded transit features with detection statistics greater than 7.1$\sigma$ were designated `Threshold Crossing Events' \citep{jenkins_overview_2010}. Although the simple transit SNR does not exactly map to their detection statistic, we use $\mathrm{SNR} \ge 7.1$ as a heuristic benchmark for whether a transit is independently detectable. 

Of the 360 sample planets, which are broadly representative of \textit{Kepler} planets overall, only 66 have an expected individual $\mathrm{SNR} \ge 7.1$. When combining all the expected transits within \textit{PLATO}'s expected two-year stare, and assuming white-noise scaling, the number of $\mathrm{SNR} \ge 7.1$ planets rises to 267. We emphasize again that the goal of this work is not to determine whether \textit{PLATO} can detect these planets independent of prior knowledge from \textit{Kepler}, but instead to assess the achievable timing precision for known systems.

\subsection{Timing performance across SNR}
\label{sec:precision_across_snr}

Panel (b) of Fig.~\ref{fig:fit_statistics} illustrates how the achievable timing precision for individual transits depends on SNR. As expected, higher SNR transits tend to yield tighter timing constraints. However, at fixed SNR there remains a substantial spread in the inferred timing precision, even for transits of the same planet, which reflects that the quality of an individual timing is stochastic in nature. A best-fit log-log regression to members of the two high-significance classes yields the relation along with the 1$\sigma$ residual scatter:

\begin{equation}
    \log_{10}{\left( \frac{\sigma_{t_0}}{t_\mathrm{duration}} \right)}=-1.1122\log_{10}{(\mathrm{SNR})}-0.3962\pm0.1578.
\end{equation}

We note that using the analytic result of \citet{carter_analytic_2008} or linear transformations thereof to predict timing precisions from SNR, ingress time, and transit duration yields a lower proportion of explained variance in the measured uncertainties than the above relation. This is likely a result of idealized noise assumptions that are not realized in the present SNR regime. Hence, we prefer our linear fit. In total, of the 3399 transit fits, 1716 fall into the prior-limited class, 594 into the low-significance class, 142 into the asymmetric significant class, and 947 into the symmetric high-significance class.

Panel (c) of Fig.~\ref{fig:fit_statistics} shows the fraction of transit fits in each classification as a function of SNR. At low SNR, most transits are either prior-limited or fail to achieve significant support over a flat model, whereas at higher SNRs the population becomes increasingly dominated by high-significance solutions. The dashed vertical lines indicate characteristic SNR values at which at least 50 per cent of transits reach a given minimum classification. Specifically, half of all transits achieve at least a low-significance but not prior limited timing constraint at $\mathrm{SNR}\simeq2.1$ (50.83 per cent of sample planets have at least this SNR), half yield at least asymmetric high-significance solutions by $\mathrm{SNR}\simeq3.8$ (33.61 per cent of planets), and symmetric high-significance timing dominates beyond $\mathrm{SNR}\simeq4.3$ (28.61 per cent of planets). These thresholds indicate that, given realistic prior knowledge from \textit{Kepler}, transits can be timed with significantly less significance than is required for independent detection.

\subsection{Validation of timing measurements}
\label{sec:timing_validation}

For each class, we investigated the relationship between the injected and extracted best-fit transit times to search for evidence of bias or underestimation of uncertainties. For each transit fit, we calculated the ratio of the difference between the injected and extracted times to the relevant uncertainty, resulting in a Z-score for each transit mid-time. Under the assumption of Gaussian posteriors, a distribution, $Z_{t_0}\sim \mathcal{N}(0,1)$ is expected.

Across all timing classes, the mean values of $Z_{t_0}$ were found to be consistent with zero to 2$\sigma$, indicating no significant bias in the recovered transit times. The dispersion of $Z_{t_0}$ depends systematically on classification: prior-limited fits exhibit sub-unity dispersion ($\sigma_{Z_{t_0}}\simeq0.7$), reflecting conservatively large uncertainties imposed by the prior boundaries, while low-significance fits are consistent with unit variance. High-significance fits show a marginally inflated dispersion ($\sigma_{Z_{t_0}}\simeq1.1$), attributable to residual non-Gaussianity and posterior asymmetry. We conclude that for all but the prior-limited fits, the extracted uncertainties reasonably reflect the true timing uncertainties.

\subsection{Performance variation with sampling cadence}

It has been shown, e.g. by \citet{Price_2014}, that finite light curve sampling leads to greater timing uncertainty than instantaneous sampling. In the preceding analysis, we analysed \textit{PLATO}'s native 25-second cadence output. Per \citet{rauer_plato_2025}, the P5 $V \le 13$~mag sample stars will predominantly receive 10-minute cadence data product, with a subset receiving 50-second cadence data, and an even smaller subset receiving 25-second data. To investigate the impact of realistic lower-frequency sampling, we repeated the fitting process of Section~\ref{sec:fitting} on the same data, but mean-average binned to 50-second and 10-minute cadence. We supersampled the \textsc{batman} fits to the original underlying 25-second cadence to account for smearing from finite integration times and to avoid introducing timing bias.

Panel (d) of Fig.~\ref{fig:fit_statistics} shows the ratio of retrieved precisions arising from data of each sampling cadence. Across the range of SNRs, we find little difference between the uncertainties in the 25- and 50-second cadence data. The 16th, 50th, and 84th percentiles of this ratio are $[0.980, 1.000, 1.021]$. Similarly, the 10-minute cadence data achieves near-identical performance to the 25-second cadence data on average, although the uncertainty ratio shows more scatter. The corresponding 16th, 50th, and 84th percentiles are $[0.947, 1.013, 1.104].$ We repeated the analysis of Section~\ref{sec:timing_validation} on the slower cadence data to search for evidence of additional uncertainty underestimation and, in both cases, found almost identical distributions of $Z_{t_0}$ to the 25-second cadence analysis. Furthermore, the number of fits falling into each classification does not deviate in either case by more than 1 per cent of the total number of fits compared to the 25-second cadence analysis. Thus, we conclude that neither 50-second nor 10-minute cadence data seriously degrades performance relative to native cadence in the studied SNR regime.

\subsection{Transit timing yield for known \textit{Kepler} systems}

A majority of the planets’ transits in our sample cannot be timed individually at a high level of significance. Under the SNR~$\ge3.8$ high-significance criterion adopted in Section~\ref{sec:precision_across_snr}, 121 of the 360 sample planets exceed the threshold for well-constrained single-transit timing measurements. This does not imply that the remaining systems are uninformative. \citet{leleu_alleviating_2021} developed a deep-learning method to detect and measure TTVs from river diagrams that remains effective for planets with low per-transit SNR, provided the TTV amplitude is sufficiently large. They demonstrate their approach on Kepler-1705~b and c, which each have a transit SNR~$\simeq1$. In our sample, 272 of the 360 planets have an individual SNR~$\ge1$.

Additionally, in multi-planet systems exhibiting TTVs, even modest additional timing constraints on one planet can improve the dynamical characterization of the entire system through coupled orbital interactions. Of the 360 planets, 189 reside in systems where at least one planet has an expected transit $\mathrm{SNR}\ge3.8$, and 332 reside in systems where at least one planet has an expected $\mathrm{SNR}\ge1$. The degree to which timing information can be recovered from lower-SNR transits depends sensitively on the amplitude and morphology of the TTV signal (see e.g. \citealt{holczer_transit_2016}) and requires system-specific dynamical modelling beyond the scope of this work. However, these results suggest that a significantly larger fraction of the planets in these systems could receive additional dynamical constraints than the fraction whose transits can be individually timed.

Another alternative to individual transit fitting is the photodynamical approach, introduced by \citet{ragozzine_value_2010}, and later applied to a variety of often multiple stellar systems, including KOI-126 \citep{carter_koi-126_2011}, Kepler-16 \citep{doyle_kepler-16_2011}, and Kepler-47 \citep{orosz_kepler-47_2012}. In contrast to the conventional two-step procedure -- in which individual transit times are first measured and subsequently fit with an independent $N$-body model -- photodynamical analyses simultaneously model the system dynamics and photometric observations within a single self-consistent model. This allows the full information content of the light curve to be utilized directly, rather than compressing the observations into a set of measured transit times.

A key advantage of the photodynamical approach is that it can remain effective even when individual transits have low signal-to-noise (see \citealt{langford_differentiable_2025}). By using the entirety of the data, information from many transits can be combined to constrain the orbital architectures and transit parameters, even when individual events are difficult to identify or characterize in isolation. Thus, the opportunity of this approach sparks further optimism that a much larger fraction of the planet sample remains tractable than their transit SNRs would initially suggest.

\subsection{Individual systems}

\citet{jontof-hutter_following_2021} used forward dynamical modelling of known \textit{Kepler} TTV systems to identify 26 systems hosting at least one planet whose projected TTVs diverge by more than 90 minutes by BJD 2460900 (corresponding to approximately 2025 August 12, 12:00 UTC), which are therefore prime targets for follow-up transit photometry. From this set, two systems were excluded from our sample because their host stars are likely subgiants (Kepler-87 and Kepler-223), and a further nine systems because none of their planets were flagged as exhibiting TTVs in the NASA Exoplanet Archive (Kepler-149, Kepler-184, Kepler-254, Kepler-266, Kepler-372, Kepler-382, Kepler-662, Kepler-822, and Kepler-1802). The remaining 15 systems, hosting 42 transiting planets, were included in our simulations. These planets, with their periods, calculated \textit{PLATO} transit SNRs, and projected TTV uncertainties at BJD 2460900, are shown in Table~\ref{tab:jontoff_planets}.

Among the 42 planets in these 15 systems, we identify five planets (KOI-1783.01, KOI-1783.02, Kepler-36~c, Kepler-177~c, and Kepler-324~c) with expected transit SNRs sufficient for individual transit timing measurements (SNR~$\ge 3.8$). These planets are distributed across four systems hosting 10 transiting planets in total. In addition, we identify 27 planets in 13 systems with expected transit SNRs~$\ge 1$); these systems collectively host 38 transiting planets. Of the 15 systems considered, only KOI-1599 and KOI-3503 host no planets with expected SNR~$\ge 1$.

The overlap between the systems identified by \citet{jontof-hutter_following_2021} as promising TTV targets and those for which \textit{PLATO} is expected to deliver useful timing information demonstrates that \textit{PLATO} has the potential to meaningfully extend the temporal baseline for a subset of known \textit{Kepler} TTV systems. Even modestly precise timing measurements, when separated by the long baseline between the \textit{Kepler} and \textit{PLATO} missions, can provide valuable information on long-term dynamical evolution.

\begin{table}
	\caption{Properties of the 42 planets in the 15 systems from the sample identified by \protect\cite{jontof-hutter_following_2021} that were included in our analysis, selected as systems in which the projected TTVs of at least one planet diverge by more than 90 minutes by BJD 2460900. Orbital periods are the default values reported in the NASA Exoplanet Archive. Transit SNRs were calculated in this work, and the projected transit-timing uncertainties for the first transit occurring after BJD 2460900 ($\delta$ TTV) are taken from \protect\cite{jontof-hutter_following_2021}. Transit SNRs exceeding the individual timing threshold of 3.8 are shown in bold.}
	\label{tab:jontoff_planets}
    \centering
	\begin{tabular}{llll}
		\hline
		\textbf{Planet} & \textbf{Period (d)} & \textbf{Transit SNR} & \textbf{$\delta$ TTV (min)}\\
		\hline
		  KOI-1599.01 & 20.44 & 0.81 & 1434 \\
KOI-1599.02 & 13.61 & 0.61 & 1132 \\
KOI-1783.01 & 134.46 & \textbf{18.96} & 95 \\
KOI-1783.02 & 284.21 & \textbf{28.76} & 562 \\
KOI-3503 b & 21.18 & 0.40 & 390 \\
KOI-3503 c & 31.83 & 0.44 & 432 \\
Kepler-29 b & 10.34 & 1.42 & 238 \\
Kepler-29 c & 13.29 & 1.34 & 310 \\
Kepler-36 b & 13.87 & 2.26 & 144 \\
Kepler-36 c & 16.22 & \textbf{13.86} & 256 \\
Kepler-49 b & 7.20 & 2.52 & 18 \\
Kepler-49 c & 10.91 & 2.41 & 33 \\
Kepler-49 d & 2.58 & 0.81 & 24 \\
Kepler-49 e & 18.60 & 1.07 & 278 \\
Kepler-54 b & 8.01 & 1.14 & 76 \\
Kepler-54 c & 12.07 & 0.42 & 131 \\
Kepler-54 d & 21.00 & 0.71 & 49 \\
Kepler-60 b & 7.13 & 1.00 & 143 \\
Kepler-60 c & 8.92 & 1.28 & 53 \\
Kepler-60 d & 11.90 & 1.47 & 230 \\
Kepler-81 b & 5.96 & 2.73 & 74 \\
Kepler-81 c & 12.04 & 2.94 & 31 \\
Kepler-81 d & 20.84 & 0.83 & 364 \\
Kepler-176 b & 5.43 & 0.47 & 18 \\
Kepler-176 c & 12.76 & 1.81 & 61 \\
Kepler-176 d & 25.75 & 1.90 & $>$1 day \\
Kepler-176 e & 51.17 & 0.71 & $>$1 day \\
Kepler-177 b & 36.86 & 3.02 & 52 \\
Kepler-177 c & 49.41 & \textbf{10.65} & 97 \\
Kepler-307 b & 10.42 & 2.06 & 10 \\
Kepler-307 c & 13.07 & 1.75 & 19 \\
Kepler-324 b & 4.39 & 0.61 & 152 \\
Kepler-324 c & 51.81 & \textbf{7.05} & 41 \\
Kepler-324 d & 34.17 & 0.84 & 218 \\
Kepler-324 e & 13.98 & 2.36 & 34 \\
Kepler-342 b & 15.17 & 2.14 & 19 \\
Kepler-342 c & 26.23 & 1.77 & 711 \\
Kepler-342 d & 39.46 & 3.08 & 28 \\
Kepler-342 e & 1.64 & 0.23 & 299 \\
Kepler-359 b & 25.56 & 0.71 & $>$1 day \\
Kepler-359 c & 57.69 & 1.21 & 759 \\
Kepler-359 d & 77.10 & 1.10 & $>$1 day \\
\hline
	\end{tabular}
\end{table}

\section{Conclusions}

In this work, we have assessed the extent to which the \textit{PLATO} mission can deliver useful transit timing measurements for known transiting exoplanets in the \textit{Kepler} field. While \textit{PLATO} will not achieve the per-transit photometric precision of \textit{Kepler} for these targets, its long-duration stare and temporal separation from the \textit{Kepler} era provide an opportunity to extend transit timing baselines by almost two decades in favourable cases.

Using end-to-end simulated \textit{PLATO} photometry, we investigated the relationship between target brightness and photometric precision, including for dim \textit{Kepler} hosts considered out of scope for the prime \textit{PLATO} mission. We found that for all but the most contaminated hosts, noise is mostly white and well modelled by the simple noise-magnitude relation of \cite{cabrera_assessment_2026}. Moreover, we found that for our sample of \textit{Kepler} hosts, \textit{PLATO} data typically exhibits 2.68--5.14 times the photometric noise of \textit{Kepler} on one-hour timescales, with the ratio varying with the number of co-pointing cameras. 

We quantified the precision with which individual mid-transit times can be recovered as a function of transit signal-to-noise ratio. We introduced a posterior-based classification scheme to distinguish between prior-limited, low-significance, and well-constrained timing measurements, and demonstrated that the transition between these regimes occurs gradually over the range $\mathrm{SNR}\simeq2.1$--4.3. Additionally, we found that timing precision is only weakly dependent on observational cadence within the range of expected data products to be delivered by the \textit{PLATO} mission. Together, we provide a framework for using catalogue-level quantities to predict photometric precision and, by extension, transit SNR. From these, the predicted mid-transit timing precision can be computed using the relation calibrated in this work.

While most individual transits fall below the threshold for independent detection, a non-negligible fraction nonetheless yield statistically powerful timing constraints when prior knowledge of the ephemeris is available. When individual transits are weak, combining multiple transits using a river diagram or photodynamical approach can provide a plausible way to recover timing information, particularly for systems with large-amplitude TTVs.

We find that \textit{PLATO} is well-positioned to contribute additional timing constraints for a subset of known TTV systems, thereby strengthening dynamical inferences through an extended temporal baseline. \textit{PLATO} can play a valuable complementary role in refining the long-term evolution of carefully selected systems identified from \textit{Kepler} data.

\section{Acknowledgements}

This work presents results from the European Space Agency (ESA) space mission PLATO. The PLATO payload, the PLATO Ground Segment and PLATO data processing are joint developments of ESA and the PLATO Mission Consortium (PMC). Funding for the PMC is provided at national levels, in particular by countries participating in the PLATO Multilateral Agreement (Austria, Belgium, Czech Republic, Denmark, France, Germany, Italy, Netherlands, Portugal, Spain, Sweden, Switzerland, Norway, and United Kingdom) and institutions from Brazil. Members of the PLATO Consortium can be found at \url{https://platomission.com/}. The ESA PLATO mission website is \url{https://www.cosmos.esa.int/plato}. We thank the teams working for PLATO for all their work.

This research has made use of the NASA Exoplanet Archive, which is operated by the California Institute of Technology, under contract with the National Aeronautics and Space Administration under the Exoplanet Exploration Program. This research has made use of the SIMBAD database, CDS, Strasbourg Astronomical Observatory, France. MAM gratefully acknowledges the support of the Science and Technology Facilities Council (STFC) through the training grant ST/X508871/1, which made this research possible. The work of RS, DR and EG has benefited from financial support by Centre National d'Etudes Spatiales (CNES) in the framework of its contribution to the PLATO mission. YNEE acknowledges support from a Science and Technology Facilities Council (STFC) studentship, grant number ST/Y509693/1. IA acknowledges the support of STFC under the CASE Industry scheme ST/W005077/1. ISL acknowledges support from UKRI grant EP/X027562/1.


\section{Data Availability}

All simulation inputs and extracted light curves may be made available upon reasonable request to the authors.



\bibliographystyle{mnras}
\bibliography{references} 





\bsp	
\label{lastpage}
\end{document}